\documentclass[aip, 
jcp, longbibliography, 
twocolumn, floatfix, superscriptaddress, reprint, 10pt]{revtex4-1}

\usepackage{cancel}
\usepackage[english]{babel}
\usepackage[utf8]{inputenc}
\usepackage[paperwidth=210mm,paperheight=297mm,centering,hmargin=2cm,vmargin=2.5cm]{geometry}
\usepackage[utf8]{inputenc}
\usepackage{verbatim}
\usepackage{comment}
\usepackage{bm}
\usepackage{amsmath}
\usepackage{amsfonts}
\usepackage{natbib}
\usepackage{pdfpages}
\usepackage{graphics}
\usepackage{float}
\usepackage{hyperref}
\usepackage[capitalize]{cleveref}
\usepackage{multirow}
\usepackage{booktabs}
\usepackage{makecell}
\usepackage{outlines}
\usepackage{lipsum}
\usepackage{siunitx}
\usepackage[version=4]{mhchem}
\usepackage{xcolor}
\usepackage{ulem}
\ifdefined\DIRACREP
\else
\newcommand{\DIRACREP}{}

\usepackage{physics}
\usepackage{xparse}
\ifdefined\COSMOMATHS
\else
\newcommand{\COSMOMATHS}{}

\newcommand{\mbf}[1]{\ensuremath{\mathbf{#1}}}

\newcommand{\oow}{\mathbf{w}^o}
\newcommand{\ooL}{\mathcal{L}^o}

\newcommand{\now}{\hat{\mathbf{w}}^o}
\newcommand{\noL}{\hat{\mathcal{L}}^o}

\newcommand{\oL}{\mathcal{L}}
\newcommand{\nL}{\hat{\mathcal{L}}}
\newcommand{\Ho}{\mathbf{H}^o}

\fi %

\NewDocumentCommand{\rep}{s d<| d|>}{%
\IfBooleanTF{#1}{
   \IfValueTF{#2}{
       \IfValueTF{#3}{\braket{#2}{#3}}{\bra{#2}}
       }{
       \IfValueTF{#3}{\ket{#3}}{}
       }
   }{
   \IfValueTF{#2}{
       \IfValueTF{#3}{\braket*{#2}{#3}}{\bra*{#2}}
       }{
       \IfValueTF{#3}{\ket*{#3}}{}
       }
   }
}

\NewDocumentCommand{\rbra}{sm}{\IfBooleanTF{#1}{\rep*<#2|}{\rep<#2|}}
\NewDocumentCommand{\rket}{sm}{\IfBooleanTF{#1}{\rep*|#2>}{\rep|#2>}}
\NewDocumentCommand{\rbraket}{smom}{
    \IfBooleanTF{#1}{
        \IfNoValueTF{#3}{\rep*<#2||#4>}{\rep*<#2|#3\rep*|#4>}
    }{
        \IfNoValueTF{#3}{\rep<#2||#4>}{\rep<#2|#3\rep|#4>}
    }
}

\NewDocumentCommand{\field}{o m e{_} e{^} o e{_} e{^}}{
\IfValueTF{#5}{\overline{
  #2\IfValueT{#3}{_#3}\IfValueT{#4}{^{\otimes #4}} %
  \otimes
  #5\IfValueT{#6}{_#6}\IfValueT{#7}{^{\otimes #7}} %
  \IfValueT{#1}{;#1}
}}{
  \IfValueTF{#4}{\overline{
     #2\IfValueT{#3}{_#3}\IfValueT{#4}{^{\otimes #4}}
     \IfValueT{#1}{;#1}
  }}
  {#2\IfValueT{#3}{_#3}}
}
}

\NewDocumentCommand{\frho}{o e{_} e{^}}{
\field[#1]{\rho}_{#2}^{#3}
}

\newcommand{\bx}{\mbf{x}}

\newcommand{\e}{a}  %

\NewDocumentCommand{\ex}{e_}{
\IfValueTF{#1}{\e_{#1}\bx_{#1}}{\e\bx}
}  %

\NewDocumentCommand{\lm}{e_}{
\IfValueTF{#1}{l_{#1}m_{#1}}{lm}
}
\NewDocumentCommand{\nlm}{e_}{
\IfValueTF{#1}{n_{#1}\lm_{#1}}{n\lm}
}
\NewDocumentCommand{\enlm}{e_}{
\IfValueTF{#1}{\e_{#1}\nlm_{#1}}{\e\nlm}
}
\NewDocumentCommand{\en}{e_}{
\IfValueTF{#1}{\e_{#1}n_{#1}}{\e n}
}
\NewDocumentCommand{\nlk}{e_}{
\IfValueTF{#1}{n_{#1}l_{#1}k_{#1}}{nlk}
}
\NewDocumentCommand{\enlk}{e_}{
\IfValueTF{#1}{\e_{#1}\nlk_{#1}}{\e\nlk}
}
\NewDocumentCommand{\enl}{e_}{
\IfValueTF{#1}{\en_{#1}l_#1}{\en l}
}

\NewDocumentCommand{\nnl}{s}{
\IfBooleanTF{#1}{n_1 n_2 l}{n_1; n_2; l}
}
\NewDocumentCommand{\ennl}{s}{
\IfBooleanTF{#1}{\en_1 \en_2 l}{\en_1; \en_2; l}
}

\NewDocumentCommand{\gslm}{s}{
\IfBooleanTF{#1}{\sigma\lambda\mu}{\sigma;\lambda\mu}
}

\fi %

\newcommand{\editor}[2]{%
  \expandafter\newcommand\csname #1note\endcsname[1]{%
    \textcolor{#2}{(\textbf{#1:} \texttt{##1})}}%
  \expandafter\newcommand\csname #1\endcsname[1]{%
    \textcolor{#2}{##1}}%
  \expandafter\newcommand\csname #1cancel\endcsname[1]{%
    \textcolor{#2}{\sout{##1}}}%
  \expandafter\newcommand\csname #1change\endcsname[2]{%
    \textcolor{#2}{\sout{##1} ##2}}%
  \newenvironment{#1text}{\color{#2}}{\color{black}}
}

\editor{FG}{red}
\editor{ray}{orange}
\editor{CBM}{magenta}

\usepackage{pdfpages}
\usepackage{pgffor}
\makeatletter
\AtBeginDocument{\let\LS@rot\@undefined}
\makeatother
\begin{document}

\setcitestyle{super}

\title{Robustness of Local Predictions in Atomistic Machine Learning Models}
\author{Sanggyu Chong}
\author{Federico Grasselli}
\author{Chiheb Ben Mahmoud}
\affiliation{Laboratory of Computational Science and Modeling, Institute of Materials, \'Ecole Polytechnique F\'ed\'erale de Lausanne, 1015 Lausanne, Switzerland}
\author{Joe D. Morrow}
\author{Volker L. Deringer}
\affiliation{Department of Chemistry, Inorganic Chemistry Laboratory, University of Oxford, Oxford OX1 3QR, United Kingdom}
\author{Michele Ceriotti}
\affiliation{Laboratory of Computational Science and Modeling, Institute of Materials, \'Ecole Polytechnique F\'ed\'erale de Lausanne, 1015 Lausanne, Switzerland}
\email{michele.ceriotti@epfl.ch}

\onecolumngrid
\begin{abstract}
Machine learning (ML) models for molecules and materials commonly rely on a decomposition of the global target quantity into local, atom-centered contributions. 
This approach is convenient from a computational perspective, enabling large-scale ML-driven simulations with a linear-scaling cost, and also allow for the identification and post-hoc interpretation of contributions from individual chemical environments and motifs to complicated macroscopic properties.
However, even though there exist practical justifications for these decompositions, only the global quantity is rigorously defined, and thus it is unclear to what extent the atomistic terms predicted by the model can be trusted. 
Here, we introduce a quantitative metric, which we call the local prediction rigidity (LPR), that allows one to assess how robust the locally decomposed predictions of ML models are. 
We investigate the dependence of LPR on the aspects of model training, particularly the composition of training dataset, for a range of different problems from simple toy models to real chemical systems.
We present strategies to systematically enhance the LPR, which can be used to improve the robustness, interpretability, and transferability of atomistic ML models.
\end{abstract}

\maketitle

\twocolumngrid

\section{Introduction}

Extensive properties of matter, such as the total energy, arise from the collective interactions between atoms and can only be rigorously defined as \emph{global} quantities that depend on the entire molecule or condensed-phase structure. Nonetheless, the last decades have seen considerable efforts toward the construction of quantum-chemical methods that exploit the quantum-mechanical nearsightedness principle\cite{prod-kohn05pnas} to perform a \emph{local} decomposition of the global quantities.\cite{yang91prl, kohn96prl, White1996, baer-head97jcp, Ochsenfeld1998, Goedecker1999} These methods either undertake physically motivated local decomposition in the calculation of a global quantity,\cite{lian+03jcp, Saravanan2003, Sodt2006, Womack2016} or perform such decomposition for the purpose of analysis.\cite{Dronskowski1993, Amadon2008} Despite the fact that the local quantities are not physical observables, such a decomposition allows one to break down the macroscopic observable for a complex structure into contributions from much simpler components, typically individiual atoms. Consequently, such methods have led to drastic improvements in the time and cost scaling of quantum-mechanical calculations, and allowed researchers to gain an enhanced understanding of the physical and chemical nature of materials.\cite{Fox2011, Lever2014, Yun2017, Skorupskii2019, Chong2020} For example, symmetry-adapted perturbation theory (SAPT) is widely used to analyze non-covalent interactions between chemical species,\cite{Szalewicz2012} and projections from delocalized plane-wave basis sets into auxiliary atomic orbitals can be used to routinely study bonding and antibonding interactions in extended systems. \cite{Maintz2016, Nelson2020, George2022}

The idea of decomposing a global quantity into contributions associated with local environments has also become a cornerstone of atomistic machine learning (ML).\cite{behl-parr07prl, bart+10prl, smit+17cs, schu+18jcp, Ko2021} By ``learning'' the global quantities as a sum of local contributions, ML models can be trained to make predictions at the local level, which are then summed up to yield the global quantity prediction for a target system. Within the context of ML, this approach has two distinct advantages. First, it makes the ML models transferable and scalable, allowing them to be easily applied to systems of vastly different length scales  (training on small cells, predicting for much larger ones),\cite{behl-parr07prl, bart+10prl} which underpins their pronounced success. Second, a local ``machine-learned'' decomposition of the global quantity can offer considerable heuristic power, because one can use the ML model to describe the complex behavior of chemical systems as resolved according to local contributions from their constituent building blocks.

Local decomposition in atomistic ML models has led to the successful development of ML interatomic potentials for a wide range of chemical systems,\cite{artr-behl12prb, soss+12prb, smit+17cs, deri-csan17prb, bart+18prx, lopa-2023} allowing researchers to access longer length and time scales in their simulations with a linear-scaling cost. It has also enabled the development of ML models for the prediction of  thermal transport in electronic insulators,\cite{Sosso2012,Verdi2021,Deng2021,Tisi2021,Pegolo2022,Brorsson2022,Langer2023} where locally predicted energies are needed in classical-like expressions of the heat flux\cite{Irving1950} used in Green--Kubo theory. In the case of ML models for the prediction of the electronic density of states,\cite{BenMahmoud2020,deri+21nature,Ellis2021,lopa+21prm,BenMahmoud2022} a meaningful correlation was found between the locally predicted density of states and the structural motifs of the environments. More recent work has shown that the ML atomic and local environment energies can be useful in interpreting the local stability of chemical environments in complex phases,\cite{schu+17ncomm, deri+18prl, bern+19acie, Cersonsky2023, Wang2023} guiding structural optimization with local information,\cite{El-Machachi2022} and can even be used as synthetic data for the pre-training of large neural network models.\cite{Gardner2023}

While the practical benefits of local decomposition for atomistic ML are clear, one must be mindful of how physically meaningful and dependable the resulting local predictions are. Since only the global quantity is rigorously defined, its decomposition into local contributions can arbitrarily take place in numerous different ways.\cite{Eckhoff2019} Furthermore, as opposed to the physically motivated decompositions typically used in quantum chemistry, the local decomposition of atomistic ML models is solely determined by the model regression, and hence it may possess an even higher degree of arbitrariness. 

In the present work, we propose a new metric, to which we refer as the {\em local prediction rigidity} (LPR), that quantifies the robustness of local predictions made by atomistic ML models. Through a series of case studies on different models, we uncover the existence of varying degrees of robustness in the local predictions, which primarily depend on the composition of dataset used for model training. We further demonstrate strategies by which the LPR can be systematically enhanced for the local environments of interest, which can ultimately improve the overall robustness, transferability, and interpretability of atomistic ML models.

\section{Theory}

Consider a generic atomistic ML model which predicts the global property $Y$ of a structure  $A$ by summing the predictions for individual atom-centered contributions, $\tilde{y}$. The task for model training is to minimize the loss function, $\oL$, which quantifies the difference between the reference values and the ML model predictions:
\begin{equation}
\oL(\mathbf{w}) = \frac{1}{2} \sum_{A \in \text{train}} \left[ Y_A - \sum_{A_i \in A} \Tilde{y}( A_i|\mathbf{w}) \right]^2 
\end{equation}
The set of optimized coefficients $\oow$ that minimizes $\oL$ is obtained by setting the derivative of $\oL$ with respect to $\mathbf{w}$ equal to 0. Close to $\oow$, one can approximate $\tilde{y}$ by a second-order Taylor expansion:
\begin{equation}
\begin{split}
    \tilde{y}(A_i|\mathbf{w}) &\approx \tilde{y}(A_i|\oow) + (\bm{\phi}^o_i)^{\sf T} (\mathbf{w} - \oow) \\
    &+ \tfrac{1}{2} (\mathbf{w} - \oow)^{\sf T} {\bm{\Psi}^o_i} (\mathbf{w} - \oow) \label{eq:ylinearization}
\end{split}
\end{equation}
where $\bm{\phi}^o_i$ is defined as
\begin{equation}
    \bm{\phi}^o_i \equiv \left.\frac{\partial \tilde{y}(A_i|\mathbf{w})}{\partial \mathbf{w}}\right|_{\oow}
\end{equation}
and $\bm{\Psi}^o_i$ as
\begin{equation}
    [\bm{\Psi}^o_i]_{ab} \equiv \left.\frac{\partial^2 \tilde{y}(A_i|\mathbf{w})}{\partial w_a \partial w_b}\right|_{\oow} \label{eq:secder}
\end{equation}
With this approximation, one can also expand the loss around $\oow$ up to the second order:
\begin{equation}
\begin{split}
    \oL(\mathbf{w}) \approx \ooL +\frac{1}{2}\, (\mathbf{w} - \oow)^{\sf T} {\Ho} (\mathbf{w} - \oow)    \label{eq:TaylorL}
\end{split}
\end{equation}
Here, $\ooL~\equiv~\oL(\oow)$, and 
\begin{equation}
\begin{split}
    \Ho &\equiv \sum_A \sum_{A_i, A_j \in A} \bm{\phi}^o_i (\bm{\phi}^o_j)^{\sf T}    \\
    &+ \sum_A \left( Y_A - \tilde{Y}_A \right) \bm{\Psi}^o_A
    \label{eq:Hessian}
\end{split}
\end{equation} 
is the Hessian of the loss evaluated at $\oow$, with $\tilde{Y}_A~\equiv~\sum_{A_i \in A} \tilde{y}( A_i|\oow)$, and $ \bm{\Psi}^o_A \equiv \sum_{A_i\in A} \bm{\Psi}^o_i$. Note that no linear term in $(\mathbf{w}-\oow)$ appears in Eq.~\eqref{eq:TaylorL} because of the optimization condition.

To assess the robustness of local predictions made by a ML model, one can consider how sensitive the model is to a change $\epsilon_k$ in a local prediction associated with an arbitrary environment $k$. 
To do so, however, explicit control over the model prediction is needed. For this purpose, one can consider the following modified loss function, which incorporates a Lagrangian term that constrains the model prediction for a local environment $k$ to be perturbed by $\epsilon_k$:
\begin{equation}
\begin{split}
    \nL(\mathbf{w}) &= \oL (\mathbf{w}) + \lambda \left[\epsilon_k - (\bm{\phi}^o_k)^{\sf T} (\mathbf{w} - \oow) \right] 
\end{split}
\end{equation}
Minimization of the new loss leads to
\begin{equation}
\begin{split}
    &\left.\frac{\partial \nL}{\partial \mathbf{w}}\right|_{\now} = \bm{0}  \quad \Rightarrow (\now - \oow) = \lambda (\Ho)^{-1}\bm{\phi}^o_k\label{eq:Deltaw}
\end{split}
\end{equation}
where $\now$ is the new array of optimal weights.
By enforcing the local prediction constraint ${\partial \nL}/{\partial \lambda} = 0$, the following expressions for $\lambda$ and $(\now - \oow)$ can be obtained:
\begin{equation}
\begin{split}
    \lambda &= \frac{\epsilon_k}{(\bm{\phi}^o_k)^{\sf T} (\Ho)^{-1}\bm{\phi}^o_k}  \\
    (\now - \oow) &= \frac{(\Ho)^{-1}\bm{\phi}^o_k}{(\bm{\phi}^o_k)^{\sf T} (\Ho)^{-1}\bm{\phi}^o_k} \epsilon_k .
\end{split}
\end{equation}
These expressions lead to algebraic simplifications resulting in the following expression for the \textit{optimized} constrained loss, where the dependence on $\epsilon_k$ is now explicitly enforced:
\begin{equation}
    \noL(\epsilon_k) \equiv \nL(\now) = \ooL + \frac{1}{2}\,\left.\frac{\partial^2 \noL}{\partial \epsilon_k^2}\right|_{\epsilon_k =0} \epsilon_k^2 . 
\end{equation}
The term
\begin{equation}
    \left.\frac{\partial^2 \noL}{\partial \epsilon_k^2}\right|_{\epsilon_k =0} \equiv \frac{1}{(\bm{\phi}^o_k)^{\sf T} (\Ho)^{-1}\bm{\phi}^o_k}
\end{equation}
is the second derivative of the constrained, optimized loss with respect to the change $\epsilon_k$ in the local prediction, and where we used $\noL(\epsilon_k =0) = \ooL$.
Note that in cases where regularization of the weights is performed, the derived expressions will only differ by the inclusion of an additional regularization term in the loss and in $\Ho$.

Ultimately, ${\partial^2 \noL}/{{\partial \epsilon_k}^2}$ describes how sensitive the model is to perturbations in a given local prediction, \emph{via} the changes in $\mathbf{w}$ caused by these perturbations. A large value of ${\partial^2 \noL}/{{\partial \epsilon_k}^2}$ indicates that the corresponding local prediction has been robustly made, as its perturbation steeply increases the loss and severely penalizes the model. Conversely, small ${\partial^2 \noL}/{{\partial \epsilon_k}^2}$ indicates that the corresponding local predictions are less robust. Since ${\partial^2 \noL}/{{\partial \epsilon_k}^2}$ essentially captures how ``rigid'' a given local prediction is, it is hereon referred to as local prediction rigidity, or LPR for short.

Having derived the LPR for a generic ML model, one can make further substitutions to obtain the expression for a specific type of model. 
For a linear model:
\begin{equation}
    \tilde{y}_k = \mathbf{x}_k \mathbf{w} \quad \Rightarrow 
    \quad
    (\boldsymbol{\phi}^o_k)^{\sf T} = \left.\frac{\partial \tilde{y}_k}{\partial \mathbf{w}}\right|_{\oow} = \mathbf{x}_k
\end{equation}
where $\mathbf{x}_k$ is a row vector containing the features of environment $k$. The Hessian reads:
\begin{equation}
\begin{split}
    \Ho &\equiv 
    \sum_A \sum_{A_i,A_j \in A} \bm{\phi}^o_i (\bm{\phi}^o_j)^{\sf T} \\ 
    &= \sum_A \sum_{A_i,A_j \in A} \mathbf{x}_i^{\sf T} \mathbf{x}_j = \mathbf{C}
\end{split}
\end{equation}
where $\mathbf{C} = \mathbf{X}^{\sf T} \mathbf{X}$ is the covariance of the feature matrix $\mathbf{X}$ of the training set, whose \textit{rows} $[\mathbf{X}]_A = \sum_{A_j \in A} \mathbf{x}_j$ are the feature vectors of the \textit{structures}. Note, also, that the second term on the right-hand side of Eq.~\eqref{eq:Hessian} vanishes since the predictions are linear in the weights.
Therefore, for the linear model:
\begin{equation}
    \mathrm{LPR}_k = \dfrac{1}{\mathbf{x}_k \mathbf{C}^{-1} \mathbf{x}_k^{\sf T}}
\end{equation}
As already mentioned, when a L$^2$ regularization with regularizer strength $\mu$ is added to the loss, it is sufficient to set $\mathbf{C} \leftarrow \mathbf{C} + \mu \mathbb{I}$.

For a sparse kernel model with L$^2$ regularization, the following expressions are obtained from direct substitution:
\begin{equation}
\begin{split}
    &\bm{\phi}^o_k = \mathbf{k}_{Mk} \\
    &\Ho = ({{\mathbf{K}_{NM}}^{\sf T}} \mathbf{K}_{NM} + \mu \, \mathbf{K}_{MM}),    
\end{split}
\end{equation}
where we adopt the notations from Ref.~\citenum{deri+21cr}, in which $N$ indicates the training set and $M$ the active set. This means that, for the sparse kernel model, the LPR of the local environment $k$ is:
\begin{equation}
    \text{LPR}_k = \frac{1}{\mathbf{k}_{Mk}^{\sf T} ({{\mathbf{K}_{NM}}^{\sf T}} \mathbf{K}_{NM} + \mu \, \mathbf{K}_{MM})^{-1}\mathbf{k}_{Mk} } 
\end{equation}
In both models, the LPR depends solely on the composition of the training set and not on the actual loss or reference energies. Such exclusive dependence on the makeup of the training set hints at the crucial importance of judiciously composing the training structures to ensure a desired level of robustness in the local predictions. 

Here, one should recognize that this property is also shared, in the context of Gaussian process regression (GPR), by estimators of the uncertainty of a prediction. For instance, in the subset of regressors (SR) approximation,\cite{rasm06book} one can express the uncertainty as:
\begin{equation}
(\Delta^2 \tilde{y}_k)_{\text{SR}} = \frac{\mu}{\text{LPR}_k}\label{eq:SRU}
\end{equation}
where, again, $\mu$ is the regularizer strength. Similar relations follow for other uncertainty estimates.\footnote{In the projected process approximation, a further term arises,
{${k(A_k, A_k)} -\mathbf{k}_{Mk}^{\sf T}(\mathbf{K}_{MM})^{-1} \mathbf{k}_{Mk}$}, which vanishes in the SR approximation due to the Nystr{\"o}m approximation for the covariance function. See Chapter 8 of Ref.~\citenum{rasm06book}.} 
It is interesting to see that in all cases, the LPR-containing term exclusively captures the dependence of $\Delta^2 \tilde{y}_k$ on the composition of the training set, as seen through the lens of the kernel used by the model.

So far, we have constructed all the main theoretical elements to quantitatively describe the robustness of a prediction for a given local environment, which in itself is not a physical observable. Here, we briefly note that in the limiting case of a structure consisting of a single type of local environment (e.g. crystalline structures in which a single Wyckoff position is occupied), the local prediction has a well-defined target of $\tilde{y}_k = Y_A/N_A$, and should therefore exhibit a maximal LPR value: any change to it would result in a change in the prediction of the global quantity of the entire structure, with a direct increase in $\oL$ that is consequential.
On the contrary, in disordered structures or structures containing atoms of different species, the local predictions would generally be far less robust and exhibit much lower LPR values due to the degeneracy in the ways in which the global quantity can be partitioned. In the following sections, we demonstrate how exactly the LPR becomes defined for the general case, and also propose strategies that can systematically improve the LPR and the robustness of local predictions made by atomistic ML models.

 \section{Proof-of-Concept Using Toy Models}

To establish and demonstrate the concepts associated with the LPR, we first construct and examine a toy model for proof-of-concept. The toy model is devised to make local predictions, $\tilde{y} \equiv \tilde{y}(x)$, depending solely on a scalar input $x$ (local features), but is trained using global targets $Y$ that are the sum of contributions from multiple $x_k$ values, i.e. $Y = \sum \tilde{y}_k$. This formulation directly corresponds to atomistic ML models, where the model predictions are made for local environments in a structure, yet regression is performed on global quantities that correspond to the entire structure. A pseudo-dataset of four data points $Y_{1, \ldots, 4}$ is constructed for training. The toy model for $\tilde{y}(x)$ is assumed to be an 8th order polynomial in $x$, and to include a L$^2$ regularization term (Figure S1).

\begin{figure}[t]
\hspace*{-0.3cm}\includegraphics[scale=1]{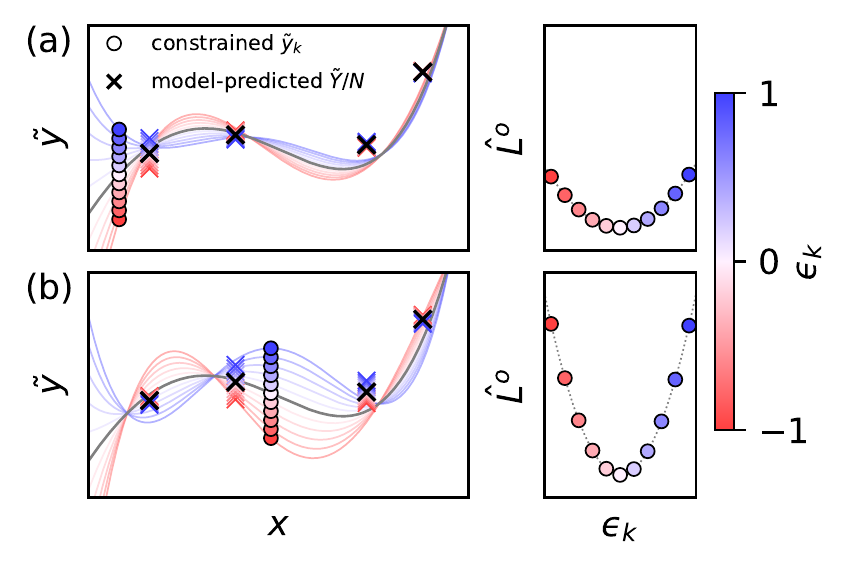}
\caption{
Graphical demonstration of the local prediction rigidity (LPR) using a numerical toy model. The left panels show, in different colors, how the model $\tilde{y}(x)$ changes when the prediction $\tilde{y}_k$ is changed by $\epsilon_k$. The prediction of the original, unconstrained model is shown in gray, and the results for the constrained models are shown in different colors that depend on $\epsilon_k$. Predictions $\tilde{Y}$ for the total target quantity are shown by crosses, and normalized by the number $N$ of elements of each group $X$ of local features. The right panels show the resulting profile of $\noL$ as dependent on $\epsilon_k$, the curvature of which corresponds to the LPR. Panels (a) and (b) report the same analysis repeated for two arbitrarily selected local features. When the $\epsilon_k$-dependent changes in $\tilde{Y}$ are small, the model re-adapts without affecting $\noL$ much and $\mathrm{LPR}_k$ is low, as shown in (a). On the contrary, if substantial changes in the total predictions $\tilde{Y}$ occur, $\noL$ is severely affected by $\epsilon_k$ and $\mathrm{LPR}_k$ is large, as presented in (b).}
    \label{fig1}
\end{figure}

As a concrete demonstration of the idea behind the LPR, we trained a series of toy models where, for a chosen $x_k$, $\tilde{y}_k$ is incrementally constrained away from the original prediction by an amount $\epsilon_k$ (right-hand side of Figure \ref{fig1}). These perturbations inevitably affect the overall optimized loss $\noL$ of the model. What ultimately results is a parabolic profile of $\noL$ around the original prediction of $\tilde{y}_k$, the curvature of which is then quantified and interpreted as the LPR. In comparing the two cases presented in the figure, one can observe the different outcomes for different choices of $x_k$: the model is far more sensitive to changes in (b) than in (a). Such higher sensitivity captures the model tendency to retain the original local prediction and corresponds to a larger LPR. Conversely, lower sensitivity is a sign of arbitrariness in the corresponding local prediction, which is associated with a smaller LPR.

\begin{figure}[t]
    \hspace*{-0.3cm}\includegraphics[scale=1]{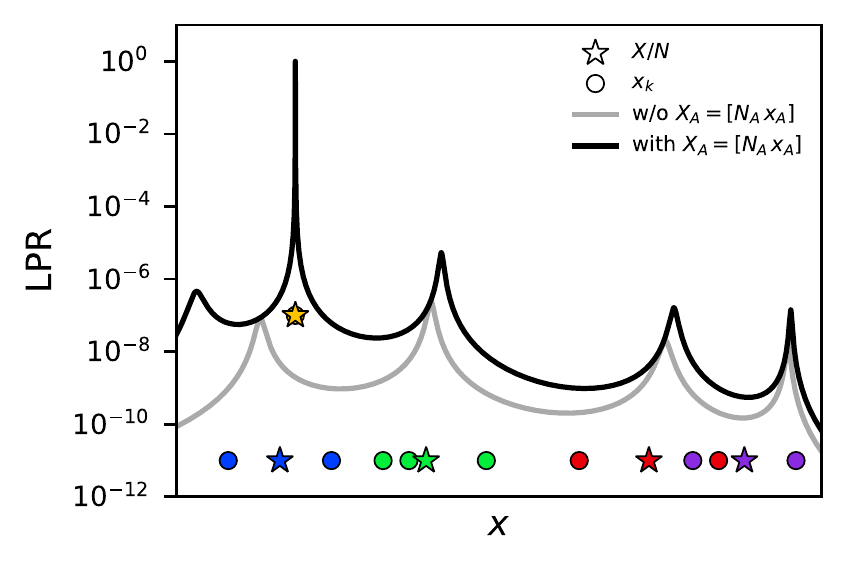}
\caption{
LPR profiles of a numerical toy model over the entire range of interest for the local feature $x$. Values of $x_k$ that appear in the training set are plotted with circles on the bottom, color-coded according to the group to which it contributes. Stars mark $X/N$ of each global data point in the training set, which corresponds to how the global quantity $\tilde{Y}$ would be predicted. The LPR profiles are shown for the model before (gray) and after (black) inclusion of a group of local features $X_A = [N_A \, x_A]$ that consists of one local feature $x_A$ replicated multiple times, shown in yellow.}
\label{fig2}
\end{figure}

Since the input value $x$ of the toy model can be continuously varied, the LPR can be computed over the entire range of interest and not only for points that are part of the training set. As shown by the gray line in Figure \ref{fig2}, this reveals the existence of peaks in the LPR profile, at which the local predictions are more robust than elsewhere. The positions of these peaks do not necessarily correspond to any particular $x_k$ found in the training set, nor to the average of the group $X_A = [x_{A_1}, x_{A_2}, \ldots ]$ of local features associated to a global quantity $Y_A$. Instead, as we will demonstrate later on, they have a delicate dependence on the degrees of freedom associated with the decomposition of the global quantity into local contributions. It is worth noting that the regularization strength $\mu$ affects the overall range of LPR and the width of the peaks that appear (Figure S2). While regularization can therefore offer some control over the robustness of local predictions, one must keep in mind that over-regularization can easily compromise the model accuracy: stable local predictions are not useful unless they lead to accurate global quantity predictions.

The analyses up to this point establish that local predictions of atomistic ML models would exhibit varying degrees of rigidity, which can be quantitatively described using the LPR. 
A subsequent question arises: what is the range of possible values for the LPR? Here, we note that the lower limit of LPR can be deduced from the expected behavior of a linear model in the data-poor, over-parameterized regime in the absence of regularization. In such a model, $\noL$ would always be 0 for any value of $\epsilon_k$, for any $x_k$ in the training set.\footnote{Here we note that there is an exception to this, which is when $x_k$ is the feature of group $X_A$ in the training set where $X_A=[N_A\,x_A]$. In real chemical systems, this corresponds to the case of single-environment structures discussed later on.} This is because the over-parameterized model would be capable of counteracting the perturbation with changes in other local predictions and always retain the correct predictions for the set of global quantities. As such, $\text{LPR}_k$ would also be 0, signifying complete arbitrariness in these local predictions.

To approach the opposite case where the LPR would instead be extremely high, we start by introducing a special class of $X_A$ made of a single input $x_A$ replicated $N_A$ times, i.e. $x_{A_1} = x_{A_2} = ... = x_A$. For such $X_A$, the local prediction $\tilde{y}_k$ is directly \emph{linked} to the global quantity, since it must target $Y_A/N_A$.  
In the context of atomistic ML, $X_A$ corresponds to what we later on refer to as a ``single-environment'' structure, where all of the local environments appearing in the structure are described by the same set of features.
For such cases, the change in $\noL$ with a perturbation in the local prediction $\tilde{y}(x_A)$ will be dramatic, since it directly affects the prediction of the global quantity $Y_A$. 
In fact, as shown by the black line in Figure \ref{fig2}, the addition to the training set of $(X_A, Y_A)$ with $X_A=[N_A\,x_A]$ creates a large peak in the LPR profile, which sits on top of $x_A$. We remark that $\text{LPR} \approx 1$ observed at the peak is not a ``hard'' limit, as there could easily be cases where inclusion of multiple $X$ groups with similar $x_k$ values or strong regularization of the model lead to LPR values that surpass 1.

\begin{figure}[t]
\hspace*{-0.3cm}\includegraphics[scale=1]{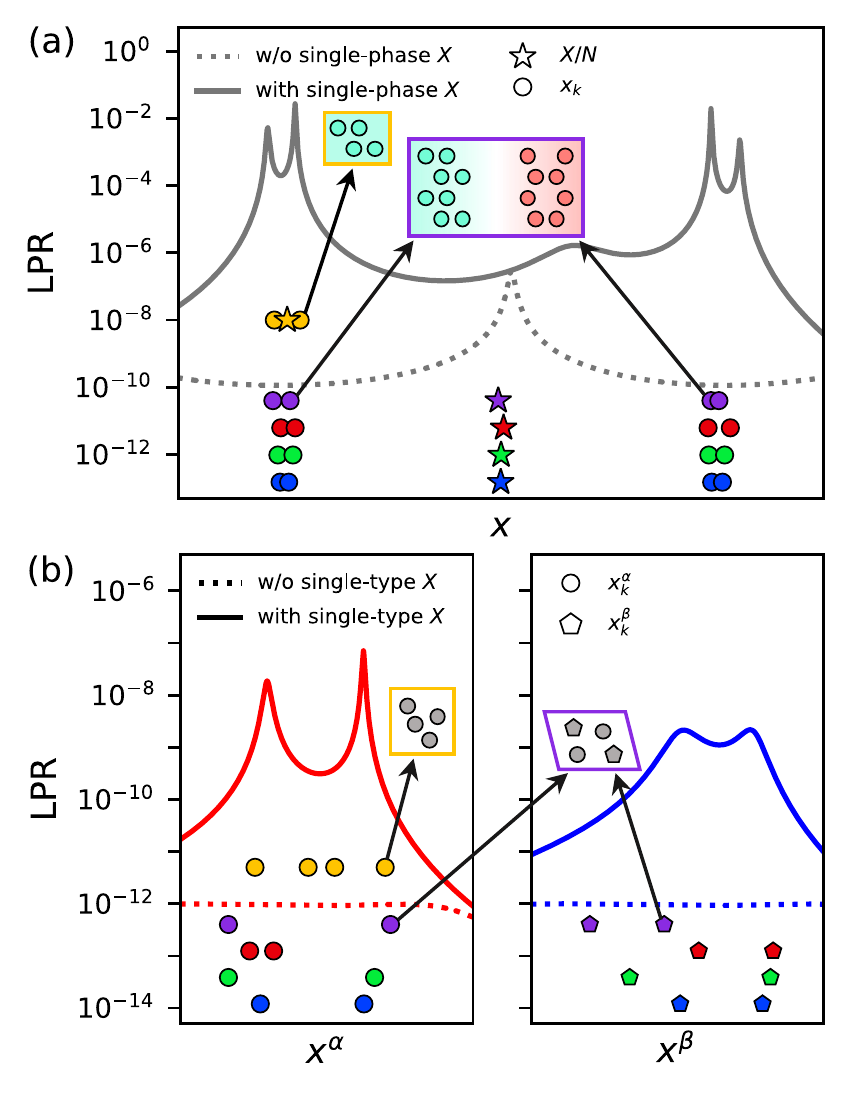}
    \caption{Effect of heterogeneity in the training data on the LPR, demonstrated using the toy model. (a) LPR profile of a model trained on a dataset composed of local feature groups $X$ with a fixed composition between two phases (dotted line), which hints at the degeneracy in the local predictions for the two phases. Inclusion of a single-phase $X$ (yellow) lifts the degeneracy and enhances the LPR \emph{for both phases} (solid line). (b) LPR profile of a composite model trained on a dataset of groups $X$ containing two distinct local feature types, $\alpha$ and $\beta$. A dataset with a fixed $\alpha$:$\beta$ compositional ratio results in very low LPR for both $\alpha$ and $\beta$ (dotted line). With the addition of $X$ only composed of $\alpha$ (yellow), the degeneracy becomes resolved and the LPR is enhanced for both (solid line).
    }
    \label{fig3}
\end{figure}

We now discuss two examples that illustrate the behavior of the LPR in more complicated scenarios. In the first example, we assume the existence of two distinct ``phases'' in the training set. This is realized by imposing a separation between two groups of local feature values, each associated with small fluctuations around one distinct value. For each $X$ in the training set, the same number of local features is sampled from the two phases. A new model is then trained, and its LPR profile is computed. The profile reveals a single peak between the two phases, which is much larger than the LPR of the actual phases (Figure \ref{fig3}a). Subsequently, another $X$ exclusively composed of local features belonging to a single phase is added to the dataset. The LPR profile of the re-trained model shows \emph{two} main peaks corresponding to the two phases, as well as an overall increase in LPR.

These differences in the LPR profile are explained by the degrees of freedom in the target quantity decomposition. Initially, partitioning of the global quantity into contributions from the two phases is completely arbitrary.
That is, the local prediction for either of the two phases can be freely made, as the prediction for the remaining phase can be adjusted to accurately recover the global quantity. The addition of a single-phase $X$ to the dataset, however, fixes the local prediction for the corresponding phase, and constrains the prediction for the remaining phase. In other words, the degeneracy in the partitioning of the global quantity into contributions from the two phases is lifted.
A similar mechanism is also at play in the second example where multi-component systems are considered, which we represent using a toy model with two distinct \emph{types} of local features, each associated with a separate prediction function.
Results in Figure \ref{fig3}b show that the effects of the previous example persist here as well, even though the predictions are made for $x_k$ values that are completely disconnected in the feature space.

Note that in both examples, there also exist further splittings of peaks in the LPR profile beyond what has been explained in terms of the phases or types. 
This suggests that similar effects must be taking place \emph{within} each phase or type, where the remaining degrees of freedom in decomposing the global quantity are further resolved. All in all, one can expect the LPR of real atomistic ML models to be determined on similar grounds, although the way in which multiple degrees of freedom are combined together, and then resolved, for structures of diverse atomic compositions would easily become quite complex.

\section{Case Studies on Demonstrative Chemical Datasets}

\begin{figure*}
    \centering  \includegraphics[scale=1]{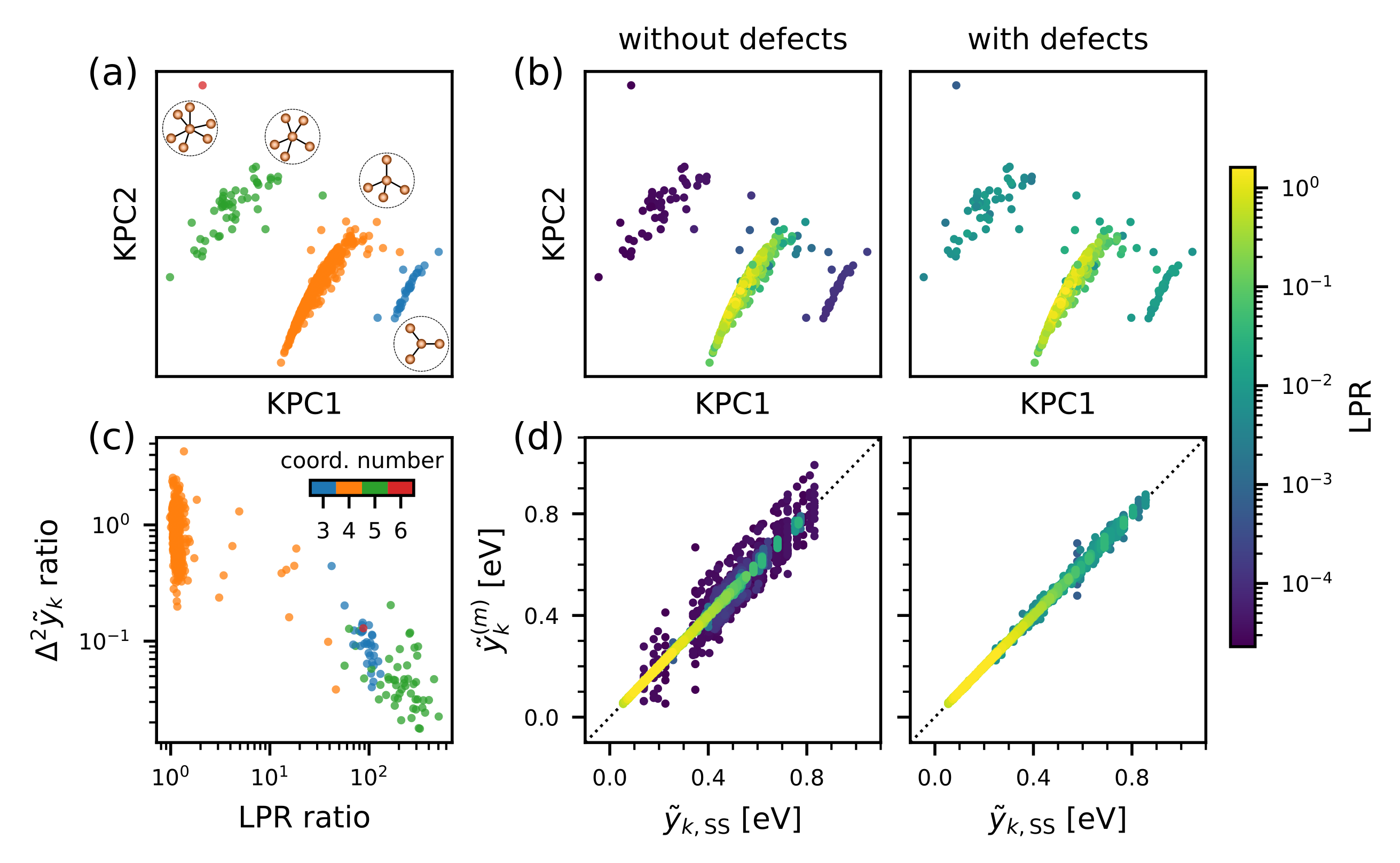}
    \caption{LPR and local energy predictions of models trained on the amorphous silicon (a-Si) dataset before and after the inclusion of structures containing under-/over-coordinated defect environments in the training set. (a) Kernel principal component analysis (KPCA) map with the points color-coded by coordination numbers of the atomic environments. For each cluster of points, a corresponding schematic environment is shown as insets. (b) KPCA map color-coded by the LPR value from each model. (c) Ratio in the variance of the committee-predicted local energies ($\Delta^2 \tilde{y}_k$) vs.\ ratio in the LPR, before and after inclusion of the defect-containing structures in the training set. (d) Parity plots of the local energies predicted by a committee of 10 models vs. the committee average prediction, where the points are color-coded by the corresponding LPR values. Energy values are reported with respect to the atomic energy of crystalline silicon. 
    }
    \label{fig4}
\end{figure*}

Having clarified the construction and the interpretation of the LPR using a toy model, we now illustrate how it can be used for actual atomistic ML models trained on chemical datasets.
For this purpose, we consider three systems: amorphous silicon (a-Si), amorphous carbon (a-C), and gallium arsenide (GaAs). In all cases, we train sparse kernel models using the total energies of the structures as the target. The predictions are made by summing the contributions from all atomic environments in a given structure. The datasets are judiciously constructed to elucidate various trends that underlie the behavior of the LPR. The atomic environments are described using the smooth overlap of atomic positions (SOAP) descriptor and kernel.\cite{bart+13prb} For demonstrative purposes, we choose hyperparameters that enhance the variation of the LPR seen in the different test cases, whilst retaining sufficient model accuracy. As we shall see in Section V, similar trends are also observed when using hyperparameters that are optimized only for the model performance. Full details of the dataset composition and ML model training are provided in the Supporting Information.

In elemental silicon under ambient conditions, each atom normally bonds with four of its neighbors to form tetrahedral coordination environments. While most environments in the a-Si dataset are close to this ideal geometry, some are ``defective'', being either under-coordinated  or over-coordinated (as detected by a bond cut-off distance of 2.7 \AA{}\footnote{The value of 2.7 \AA{} is smaller than the commonly used value of 2.85 \AA{} for silicon bond detection. This value was heuristically chosen to prevent the neighboring silicon atoms from being present in distance range of the smooth cutoff function of the SOAP descriptor.}). 
We study the effect of including defect-containing structures in the training set on the resulting LPR of the model. For analysis, kernel principal component analysis (KPCA) is performed to plot the local environments in a low-dimensional representation of the feature space, and then color-coded by the LPR to study the trends.
Figure \ref{fig4} shows that the LPR of under/over-coordinated environments in the test set is comparatively low for a ML model trained without any defect-containing structures in the training set. 
When 10\% of the training set is replaced by the defect-containing structures, the LPR of the defect environments is enhanced by several orders of magnitude. 
The variance of local energy predictions across a committee of models (herein referred to as $\Delta^2 \tilde{y}_k$ without any subscripts) significantly decreases for the defective environments, in line with the link between the LPR and GPR uncertainty.
This is further corroborated by the change in $(\Delta^2 \tilde{y}_k)_{\mathrm{SR}}$, which is reduced by up to 112 meV (compared to 3--6 meV RMSE per atom for a test set of defect-containing structures). 

\begin{figure}
   \hspace*{-0.3cm}\includegraphics[scale=1]{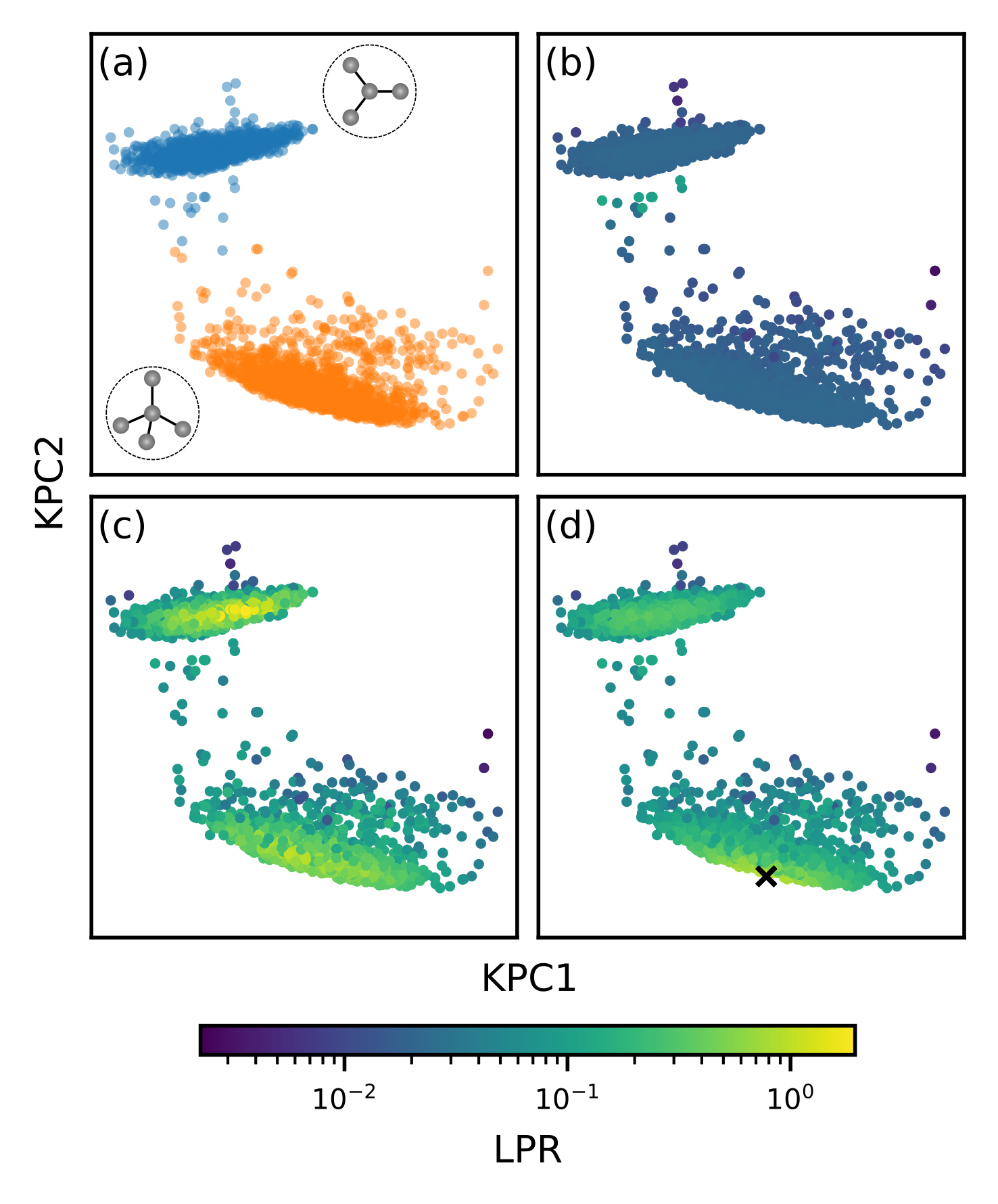}
    \caption{KPCA maps for an ensemble of amorphous carbon structures colored by the hybridization of the atoms, shown in panel (a), and then by the LPR of the models trained on differently composed training sets. The top and bottom clusters of points correspond to sp$^2$ and sp$^3$ environments, respectively, and the corresponding schematic environments are shown as insets. Panel (b) shows the results obtained when the model is trained on a dataset exclusively composed of structures that retain 1:1 ratio between sp$^2$ and sp$^3$ carbons. Panel (c) shows the case where 10\% of the dataset is replaced with structures exhibiting a different sp$^2$ to sp$^3$ ratio. Panel (d) shows the case when a single structure in the dataset is replaced with the crystalline diamond structure, for which the location in the KPCA map is marked with a cross.}
    \label{fig5}
\end{figure}

The a-C dataset is comprised of structures containing a mixture of ``sp$^2$'' and ``sp$^3$'' carbon environments (defined by counting bonded neighbors up to a cut-off distance of 1.82 \AA{}\footnote{Similar to silicon, the smaller value of 1.82 \AA{} as opposed to 1.85 \AA{} was heuristically chosen to prevent the neighboring carbon atoms from being present in distance range of the smooth cutoff function of the SOAP descriptor.}). 
This effectively introduces a degree of freedom in the decomposition of the total energy into the contributions from the two distinct types of carbon environments.
In fact, when the model is trained on a dataset exclusively composed of structures with a 1:1 ratio between sp$^2$ and sp$^3$ carbons, energy partitioning between the two carbon types is performed rather arbitrarily, as evident from the LPR (Figure \ref{fig5}b).
Drawing on what was previously observed for the toy model on an artificial two-phase system (Figure \ref{fig3}a), we introduce structures that exhibit a different ratio between the two carbon types into the training set to lift the apparent degeneracy. 
Indeed, Figure \ref{fig5}c shows that when 10\% of the training set is replaced by structures with a different ratio between sp$^2$ and sp$^3$ carbons, the LPR increases for \textit{both}. 
The increased robustness in local energy predictions is confirmed by a notable decrease in $\Delta^2 \tilde{y}_k$ (Figure S3).

Another effect that can be demonstrated with the a-C dataset is the enhancement of LPR from the inclusion of ``single-environment'' structures. Both sp$^2$ and sp$^3$ carbons have crystalline analogues, graphite and diamond, where the local energy is uniquely defined as all of the atoms in the structure are equivalently described with the same set of local features. In Figure \ref{fig5}d, it is shown that the LPR improves significantly when a single diamond structure is included into the training set, especially for the sp$^3$ environments that are close to diamond on the kernel PCA map. Inclusion of the diamond structure is also capable of resolving the energy decomposition degeneracy between the sp$^2$ and sp$^3$ carbon atoms, and hence improvement in the LPR is observed for the sp$^2$ environments as well. Once again, this can be equivalently seen as the decrease of $\Delta^2 \tilde{y}_k$ for both sp$^2$ and sp$^3$ environments (Figure S4). These results emphasize the importance of recognizing and resolving degeneracies associated with distinct phases or atomic types in a dataset, which could be as simple as including a small number of single-environment structures associated with each phase/type.

\begin{figure}
    \hspace*{-0.3cm}\includegraphics[scale=1]{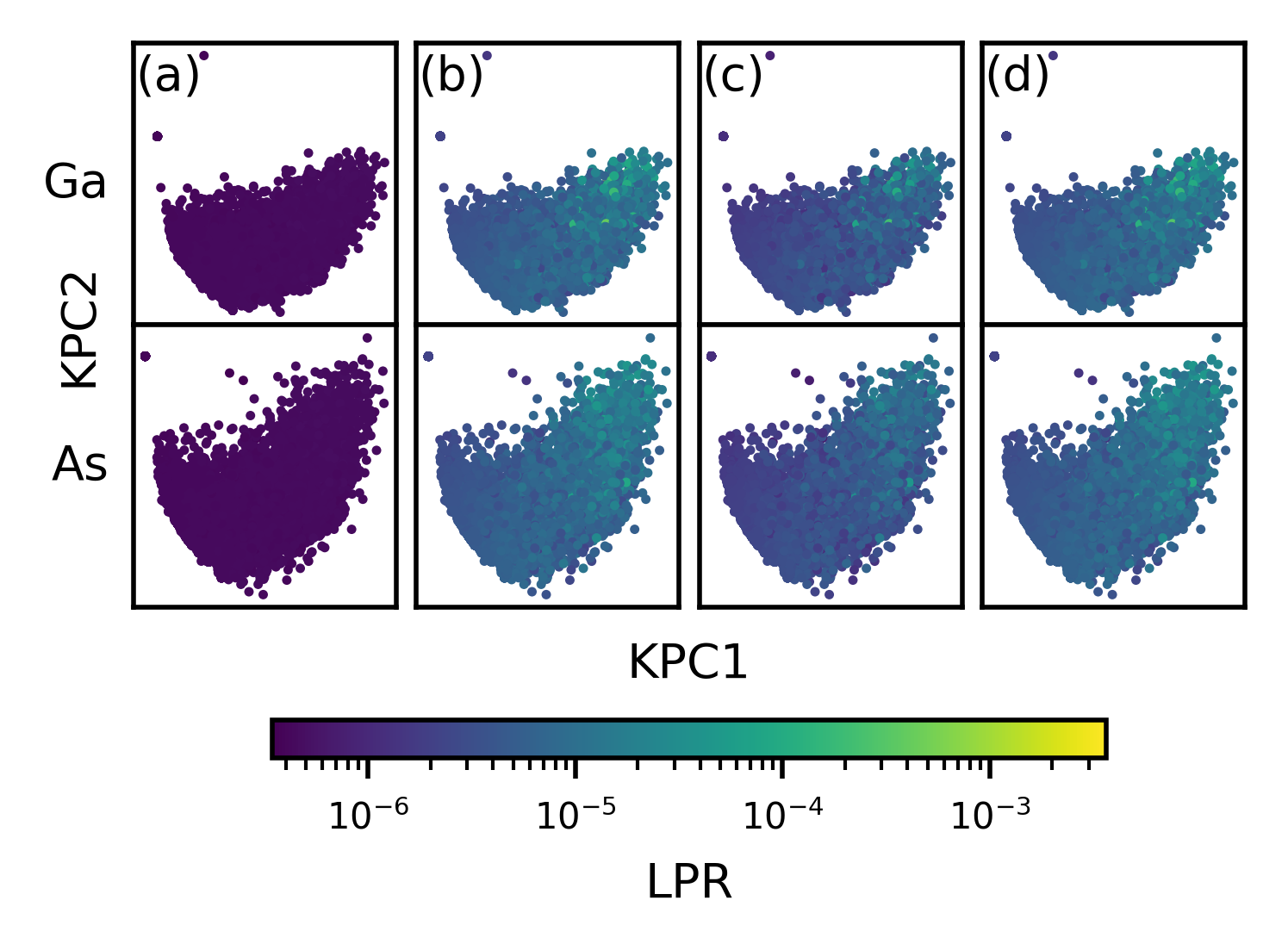}
    \caption{KPCA maps for a GaAs dataset. Separate maps are shown for Ga (top row) and As (bottom row) atomic environments, and the points are color-coded by the corresponding LPR values. Results are shown for a series of models trained on datasets with different compositions: 
    (a) exclusively composed of structures with a Ga:As ratio of 1:1;
    (b) with 10\% of the dataset replaced with structures exhibiting a different Ga:As ratio; 
    (c--d) with 10\% of the dataset replaced with pure Ga or pure As structures, respectively.
    }
    \label{fig6}
\end{figure}

Finally, effects in the LPR associated with the presence of multiple atomic species in the structures are explored using a GaAs dataset, a physical analogue of the toy model presented in Figure \ref{fig3}b. For a model trained exclusively on the structures with 1:1 stoichiometric composition (Figure \ref{fig6}a), the LPR remains consistently low for both Ga and As, and does not even show significant variations in the values within. This signifies close-to-complete arbitrariness in the energy decomposition between the two species. Figures \ref{fig6}b--d show the results when 10\% of the training set is replaced by structures with a different Ga:As ratio, pure Ga structures, or pure As structures, respectively. In all cases, the degeneracy in the local energy decomposition is resolved, the LPR of both Ga and As is notably enhanced, and $\Delta^2 \tilde{y}_k$ becomes significantly smaller (Figures S5--7). 

These case studies demonstrate that, similar to what was previously observed for the toy model, robustness in the local predictions can drastically vary even for atomistic ML models trained on real chemical systems, and the degree of robustness quantified by the LPR depends considerably on the composition of the training set. To improve the LPR and hence the robustness of the local predictions, one must first ensure sufficient representation of all local environments of interest in the training set structures. In the case of chemical systems with distinct phases/local environments, or species, the training set should be carefully composed so that the degeneracy in the energy decomposition could be resolved as much as possible. We note in closing that these effects are not specific to the sparse kernel model, as similar trends are consistently observed when the analyses are repeated for linear ridge regression models (Supporting Information).

\begin{figure*}
    \centering
    \includegraphics[scale=0.6]{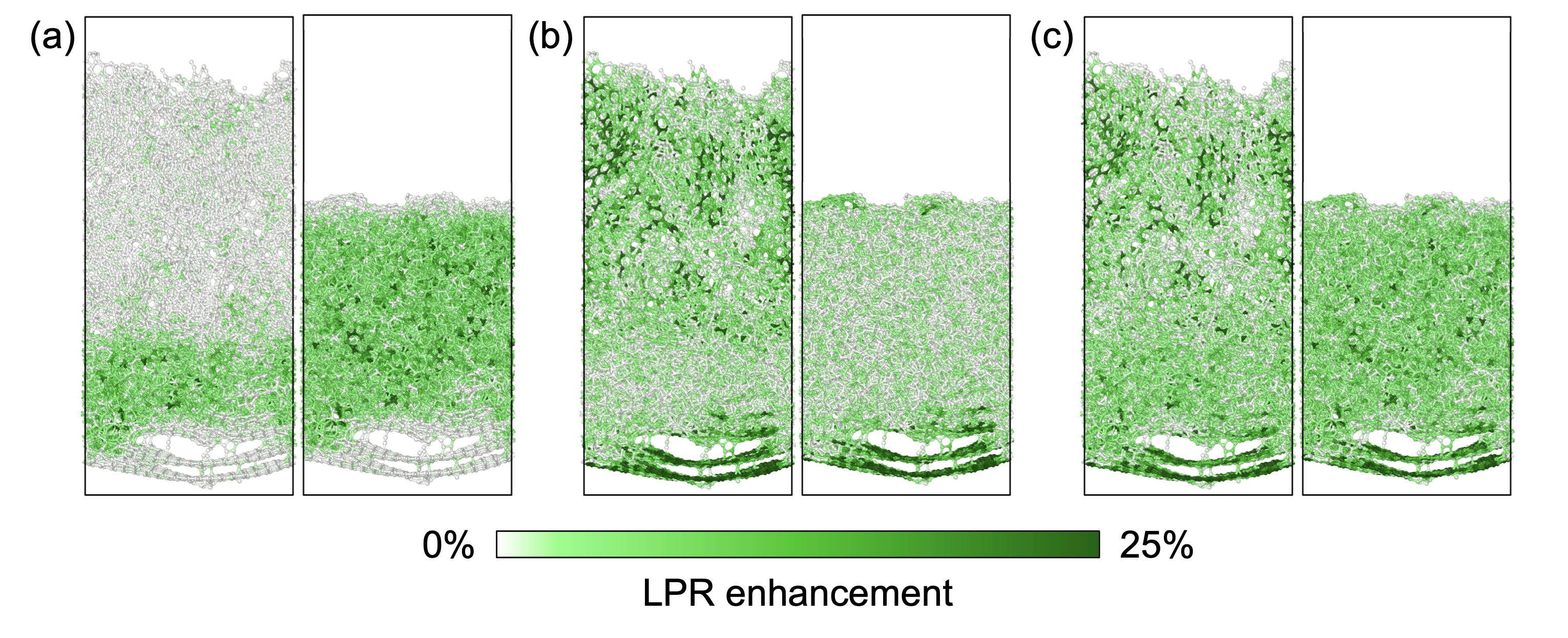}
    \caption{Enhancement in the LPR for low-density (left) and high-density (right) carbon films taekn from Ref.~\citenum{caro+20prb} with the inclusion of single-environment structures in the training set. Results are shown for SOAP-based sparse kernel models of elemental carbon as described in the text. In all cases, the enhancement is computed with respect to a baseline model trained on 1000 amorphous carbon structures. (a) LPR enhancement when 10 training set structures are replaced with diamond-like single-environment structures, which constitutes only 1\% of the training set. Enhancement is mostly observed for the local environments of the high density film. (b) LPR enhancement when 10 structures are replaced with graphite-like single-environment structures. Enhancement takes place for the environments found in the low density film. (c) LPR enhancement when 5 diamond-like and 5 graphite-like structures are incorporated into the training set. Enhancement is consistently observed for the local environments of both films.
    Structures were visualized using OVITO.\cite{Stukowski2010}
    }\label{fig7}
\end{figure*}

\section{A Realistic Application}

The demonstrative case studies of Section IV elucidate the existence of varying degrees of robustness in the local predictions made by atomistic ML models as quantified by the LPR, and how it depends on the composition of the training set. 
In this section, we further expand upon our findings to devise strategies to systematically enhance the LPR and the robustness of the local predictions. In the general case, the degeneracy in the local decomposition is expected be far more complex than those seen in the previous case studies. One failsafe strategy to guarantee high LPR would be to judiciously compose the training set, from scratch, in a manner that resolves the degeneracy for as many local environments of interest as possible. In most cases, however, such an approach would be hindered by data availability and computational cost associated with generating the necessary new data.

Here, we instead propose the generation and inclusion of single-environment structures into the training set as a simple yet effective strategy in which the LPR can be systemically enhanced. As previously discussed, single-environment structures are those composed of one local environment replicated multiple times, as the case of single-species crystalline structures with a single Wyckoff position, which leads to a unique definition in the local prediction target. This results in a maximal LPR value for the corresponding local environment, and heightened LPR for sufficiently similar environments around it (Figures \ref{fig2} and \ref{fig5}d). Then, by introducing single-environment structures that closely resemble the local environments of interest to the training set, one can induce a notable enhancement in the LPR and improve the robustness of the model predictions. One should also note that due to their high symmetry (i.e., small and simple unit cells), generating such structures and obtaining their reference data is considerably cheaper than constructing the rest of the dataset.

To demonstrate this strategy, we present a realistic case study where the inclusion of single-environment structures in the model training enhances the LPR for the local environments of interest in the target system. For this, we direct our attention to the studies of a-C films conducted by Caro et al.\cite{Caro2018, caro+20prb} Benefiting from the scalability of atomistic ML models, the authors carried out large-scale simulations to uncover the precise growth mechanism of a-C films when they are grown by the deposition of highly energetic ions onto a substrate. They also computed the GPR-based error estimates to ensure that the uncertainty in the model predictions remains reasonably low throughout their simulations. Here, we further expand on this by showing that it is possible to systematically enhance the LPR for particular local environments of interest and reduce the uncertainty in the model predictions.

The a-C films from Ref.~\citenum{caro+20prb} significantly vary in their mass densities, depending on the energies of incident atoms for deposition. The films hence exhibit different similarities in their local environments to graphite (lower density) or diamond (higher density), which are both crystalline, single-environment structures. 
As such, we train and analyze carbon ML models before and after the inclusion of single-environment structures obtained as high symmetry distortions of diamond or graphite. 
First, we train a SOAP-based sparse kernel model with the identical set of hyperparameters used in the reference study,\cite{deri-csan17prb} on 1000 randomly chosen a-C structures from the authors' published dataset.
The model is subsequently re-trained under the same conditions, but with 10 structures in the training set replaced with diamond and/or graphite and derivative structures. The derivative single-environment structures are generated by distorting the unit cell vectors whilst occupying the original, single Wyckoff position (Figure S16). This procedure ensures that while the local environment changes, all atoms in the unit cell are still described equivalently. Full details of the model training and derivative single-environment structure generation are provided in the Supporting Information.

Figure \ref{fig7} shows the enhancement in the LPR with the inclusion of single-environment structures for the representative low density and high density a-C films. 
When 10 diamond-like single-environment structures are included, the LPR enhancement is mostly observed for the local environments in the high density a-C film (Figure \ref{fig7}a, stronger green color). 
Conversely, when 10 graphite-like single-environment structures are included, the LPR enhancement takes place primarily for the environments in the low density film (Figure \ref{fig7}b). 
For 100 local environments across both films that are the most similar to the newly added single-environment structures, we observe an average LPR enhancement of 31\% for the diamond-like environments, and 54\% for the graphite-like environments. 
Interestingly, when both types of single-environment structures are incorporated into the training set, i.e., 5 diamond-like and 5 graphite-like single-environment structures, enhancement of the LPR is observed throughout both low and high density a-C films (Figure \ref{fig7}c), with an average enhancement of 36\% for the 200 previously selected local environments. 
In terms of $(\Delta^2 \tilde{y}_k)_{\mathrm{SR}}$, inclusion of the single-environment structures reduces the uncertainty by up to 87\%.
Such improvements take place whilst the accuracy of the models remain largely the same, where the \%RMSE on the test set changes from 12\% to 14\% at most.

These results prove that generation and inclusion of single-environment structures similar to the local environments of interest is a highly effective strategy to systematically enhance the LPR and improve the robustness in the local predictions of the ML model. It is striking to see that notable enhancement is already induced by replacing only 1\% of the dataset with single-environment structures. While only diamond- and graphite-like single-environment structures are considered here, the discovery and inclusion of other single-environment samples, diverse in their structures yet similar to the local environments of interest, might likely induce further enhancements in the LPR.

\section{Extension to Neural Network Models}

\begin{figure}
    \hspace*{-0.3cm}\includegraphics[scale=1]{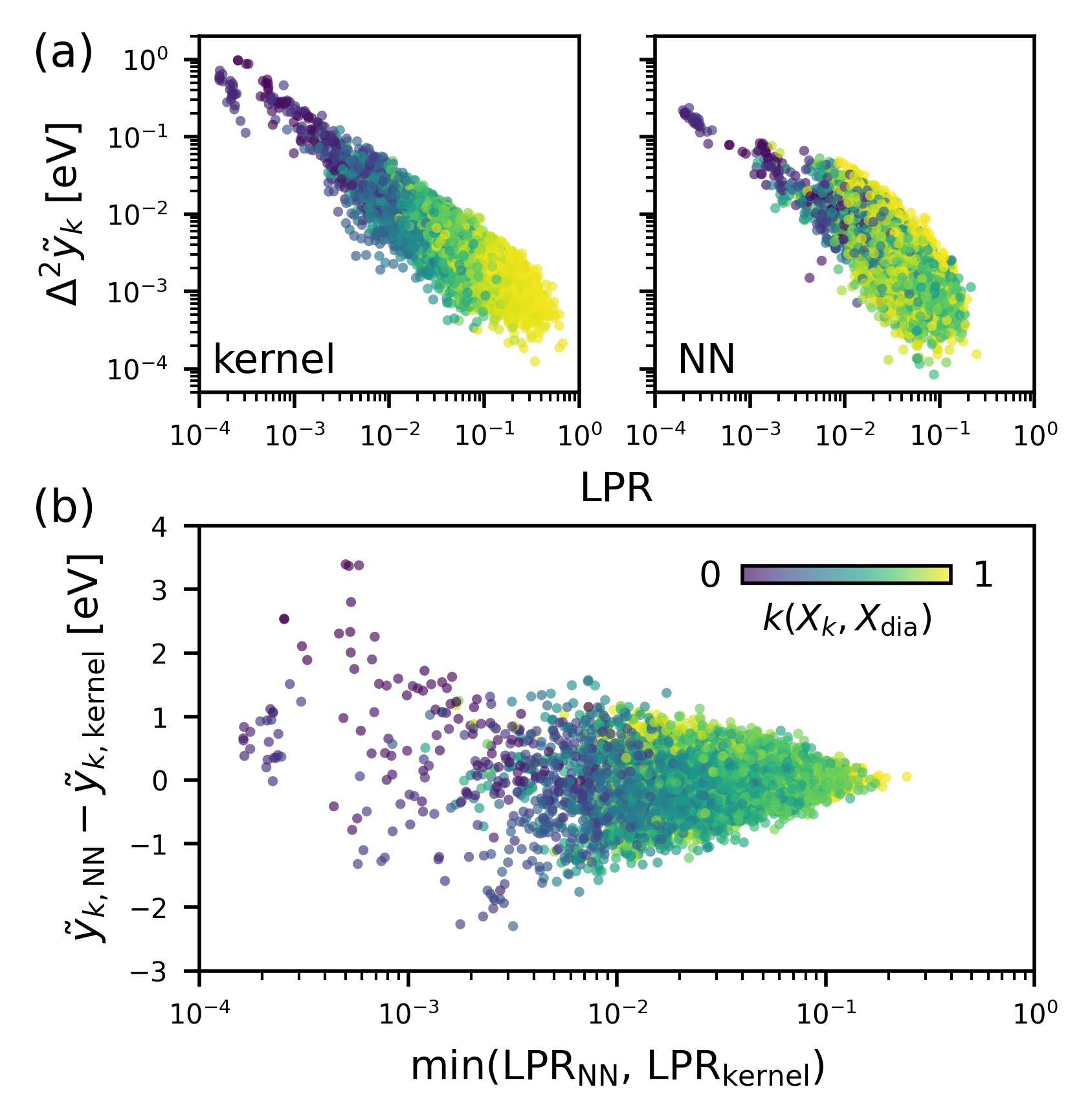}
    \caption{Extension of the LPR analysis to the neural network (NN) model. (a) $\Delta^2 \tilde{y}_k$ for a committee of 10 models vs. the LPR, calculated for the low density carbon film using the sparse kernel model (left) and NN model (right). (b) Difference in the local energy predictions vs. minimum LPR between the sparse kernel model and the NN model. In all cases, data points are colored by the SOAP kernel similarity of the local environments to pristine diamond.
    }
    \label{fig8}
\end{figure}

Thus far, we have applied the LPR analysis only to linear and kernel models, which are associated with a convex loss function that can be minimized analytically. We now extend our study to the case of neural network (NN) models. 
NNs are a large class of regression methods in atomistic ML.\cite{behl-parr07prl, smit+17cs, schu+18jcp, Batatia2022, Batzner2022, Ko2021, Musaelian2023, pozdnyakov2023smooth} They are generally regarded to be far more ``flexible'' than their linear counterparts, given the significantly larger number of weight parameters involved in training the model. 
One important detail that sets NN models apart from the previously mentioned ones is that they cannot be optimized in an analytical, deterministic way: model training is often carried out with recursive numerical methods and does not exactly reach the actual minimum, which is an assumption underlying the formulation of the LPR. 
Here, we assume that the NN models trained for our analysis are close enough to the minimum for the LPR formulations to still be applicable. 
Another point to note is that the second-order derivatives $\bm{\Psi}^o_A$ of Eq.\ \eqref{eq:secder} does not vanish in general for NN models. 
Nevertheless, as is customary in the context of nonlinear optimization,\cite{Press2007-LevMar} we assume a negligible statistical correlation between $(Y_A - \tilde{Y}_A)$ and $\bm{\Psi}^o_A$ over the training set and drop the second term on the right-hand side of Eq.\ \eqref{eq:Hessian}. In practice, we obtain $\Ho$ by computing and accumulating $\bm{\phi}^o_i$ for the local environments in the training set by using the automatic differentiation framework in PyTorch.\cite{pytorch}

We train a simple multi-layer perceptron model with 2 hidden layers, each composed of 16 nodes with a nonlinear sigmoid activation function. The model is trained on the same carbon dataset as in the previous section with the SOAP power-spectrum vectors as the input layer, and their local energies predicted at the output layer. 
We adopt the Behler--Parrinello approach of summing the local NN predictions outside of the NN model to regress global quantities.\cite{behl-parr07prl} 
We also perform explicit L$^2$ regularization of the NN model weights, rather than the conventional early stopping with respect to a validation set, to retain the loss function used in deriving the LPR and ensure comparability with the previous linear models. Full details of NN model training and LPR calculation are provided in the Supporting Information. The test set \%RMSE for the NN model is 12\%.

For the analysis, the LPR and $\Delta^2 \tilde{y}_k$ of the low density carbon film from the previous Section are calculated for the sparse kernel model and the NN model. In Figure \ref{fig8}a, both models exhibit a clear inverse proportionality between the LPR and $\Delta^2 \tilde{y}_k$ for the local predictions. This emphasizes the clear relationship between the LPR and the uncertainty in the local predictions and how it also extends to nonlinear, NN models. Additionally, in Figure \ref{fig8}b, difference in the local energy predictions diminishes as the minimum LPR increases between the two models. This shows that the LPR accurately captures how much the two models agree in their local predictions, which supports our interpretation of the LPR as a measure of how trustworthy the local predictions of the ML models are.

An interesting difference to be noted here is the correlation between the LPR (or $\Delta^2 \tilde{y}_k$) and the local environment similarity to diamond. In the sparse kernel model, $\Delta^2 \tilde{y}_k$ decreases with increasing similarity to diamond, which stems from the abundance of diamond-like environments in the training set (Figure S17). For the NN model, such a correlation is absent, and the lowest values of $\Delta^2 \tilde{y}_k$ are also observed for the environments that notably differ from diamond. This suggests that the heuristic observations of the dependence of the LPR on the dataset composition, which we have seen for linear and kernel models, applies only partially to the nonlinear NN model. Given the vast variety of NN architectures and algorithms for atomistic ML, further work is required to better understand how exactly the NN architecture and training details affect the LPR of the resulting NN models.

\section{Conclusions}

While the local decomposition approach commonly adopted by atomistic ML models has proven to be very successful, it inevitably introduces a degree of arbitrariness in the model predictions, which are made locally and without a well-defined target. 
To understand how meaningful these local predictions can be, we have devised the local prediction rigidity (LPR),  which quantifies the robustness of the local predictions made by the atomistic ML models. 
For a range of models and datasets, we demonstrated that the LPR can vary drastically between different local environments. Local predictions of atomistic ML models should therefore be interpreted cautiously, and the LPR should be taken into consideration alongside the model predictions.

Our analyses have also shown that the process in which the LPR becomes determined for a ML model prediction is largely dependent on the degeneracies associated with the local decomposition of the target global quantities. To systematically improve the LPR, the dataset for model training should be judiciously constructed to eliminate as much of the degeneracy as possible. For this, all local environments of interest should be sufficiently well-represented in the dataset for model training. In cases where multiple atomic types or species are present, many different chemical and structural compositions must be probed by the dataset to eliminate the degeneracy between the types or species. One can also generate and include single-environment structures to systematically enhance the LPR of a model for the local environments of particular interest. Lastly, the LPR can even be utilized as a metric of uncertainty across different types of atomistic ML models.

The clear connection between the LPR and uncertainty suggests that measures of error in the local predictions, which are readily available in several widely used models, can be used to compute a substitute for the LPR. This makes it possible for one to easily expand on the insights found in our study for a wider range of atomistic ML models.
As the derivation of LPR is not limited to the atomic decomposition primarily dealt with in this study, it can be extended to other decomposition schemes: multiple body-order decomposition, short-range versus long-range decomposition, and so forth.
This allows one to precisely identify where the ML model lacks robustness in the predictions, and to propose effective ways to improve it.

\section*{Data Availability}
The data that support the findings of this study are available in the paper and the Supporting Information. Data analysis scripts employed to generate the plots and relevant results within this paper are available on the Materials Cloud platform.\cite{tarl+20sd} See DOI: \href{https://10.24435/materialscloud:re-0d}{10.24435/materialscloud:re-0d}

\section*{Acknowledgements}

We thank Filippo Bigi for reviewing the derivation of the LPR in the context of neural network models. S.C., F.G., and M.C. acknowledge funding from the European Research Council (ERC) under the European Union’s Horizon 2020 research and innovation programme Grant No. 101001890-FIAMMA. F.G. also acknowledges funding from the European Union's Horizon 2020 research and innovation programme under the Marie Sk\l{}odowska-Curie Action IF-EF-ST, grant agreement number 101018557 (TRANQUIL).
J.D.M.\ acknowledges funding from the EPSRC Centre for Doctoral Training in Inorganic Chemistry for Future Manufacturing (OxICFM), EP/S023828/1.

\appendix 

\onecolumngrid
\clearpage
\newpage

\section*{Table of Contents Graphic}
\begin{figure}[!h]
\centering
\includegraphics[width=0.8\textwidth]{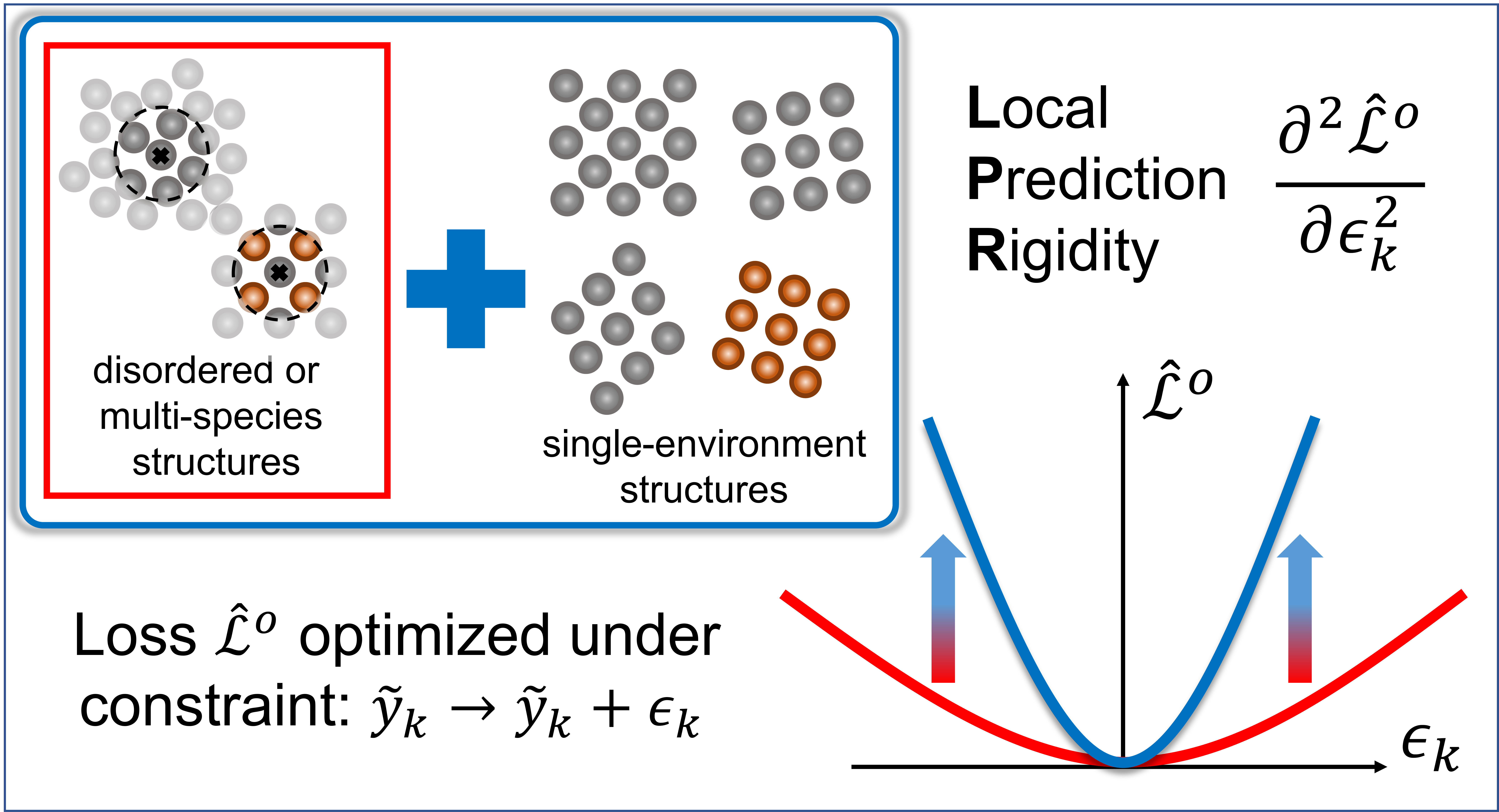}
\end{figure}
\end{document}